\documentclass[conference]{IEEEtran}
\IEEEoverridecommandlockouts
\usepackage{cite}
\usepackage{amsmath,amssymb,amsfonts}
\usepackage{algorithmic}
\usepackage{algorithm2e}
\usepackage{graphicx}
\usepackage{textcomp}
\usepackage[T1]{fontenc}
\usepackage{xcolor}
\usepackage{hyperref}
\graphicspath{{./img/}}
\newcommand{\squeezeup}{\vspace{-2mm}}
\def\BibTeX{{\rm B\kern-.05em{\sc i\kern-.025em b}\kern-.08em
    T\kern-.1667em\lower.7ex\hbox{E}\kern-.125emX}}
\begin{document}

\title{CoronaVis: A Real-time COVID-19 Tweets Data Analyzer and Data Repository}

\author{\IEEEauthorblockN{
Md. Yasin Kabir}
\IEEEauthorblockA{\textit{Department of Computer Science} \\
\textit{Missouri University of} \\
\textit{Science and Technology, USA}\\
mkabir@mst.edu}
\and
\IEEEauthorblockN{
Sanjay Madria}
\IEEEauthorblockA{\textit{Department of Computer Science} \\
\textit{Missouri University of} \\
\textit{Science and Technology, USA}\\
madrias@mst.edu}
}

\maketitle

\begin{abstract}
Due to the nature of the data and public interaction, twitter is becoming more and more useful to understand and model various events. The goal of CoronaVis is to use tweets as the information shared by the people to visualize topic modeling, study subjectivity and to model the human emotions during the COVID-19 pandemic. The main objective is to explore the psychology and behavior of the societies at large which can assist in managing the economic and social crisis during the ongoing pandemic as well as the after-effects of it. The novel coronavirus (COVID-19) pandemic forced people to stay at home to reduce the spread of the virus by maintaining the social distancing. However, social media is keeping people connected both locally and globally. People are sharing information (e.g. personal opinions, some facts, news, status, etc.) on social media platforms which can be helpful to understand the various public behavior such as emotions, sentiments, and mobility during the ongoing pandemic. In this work, we develop a live application to observe the tweets on COVID-19 generated from the USA. In this paper, we have generated various data analytics over a period of time to study the changes in topics, subjectivity, and human emotions. We  also share a cleaned and processed dataset named CoronaVis Twitter dataset (focused on United States) available to the research community at \url{https://github.com/mykabir/COVID19}. This will enable the community to find more useful insights and create different applications and models to fight with COVID-19 pandemic and the future pandemics as well. 
\end{abstract}


\section{Introduction}

At the time of writing this document, there were more than 11 million confirmed cases of novel corona virus cases all over the world, and around 3M people are infected\footnote{\url{https://coronavirus.jhu.edu/map.html}} in USA alone. The number of total fatal cases exceeded 500k globally. The number of infected people, active cases, and fatality keep rising every day. The first confirmed case of novel corona virus disease was reported in Wuhan, China. However, over the last 5 months, the virus spread explosively all over the world. At the time of writing this document, the United States has the maximum number of corona virus cases and fatalities trailing by Brazil, and then followed by India.

Every country is taking preventive measurements to fight against the COVID-19 pandemic. Social distancing or stay-at-home became the most widely used directive all over the world. Social distancing is forcing people to stay at home, and as a result, it is impacting the public event, business, education, and almost every other activity associated with the human life. People are also losing their jobs, and getting infected from corona and thus, stress is rising at the personal and at the community levels. Studies of behavioral economics show that emotions (Joy, Anger, Worry, Disgust, Fear, etc.) can deeply affect the individual behavior and decision-making. 

Social networks have the hidden potential to reveal valuable insights on human emotions at the personal and community level. Monitoring tweets could be valuable particularly during and after COVID-19 pandemic as the situation and people reaction both are changing every moment during this unpredictable time. Thus, the analysis of twitter data might play a crucial role to understand the people behavior and response during the COVID-19 pandemic. Recent works \cite{chen2015crime, gerber2014predicting, grover2019polarization, kabir2019deep} show that twitter data, and human emotions analysis can be useful for predicting crimes, stock market, election vote, disaster management, and more. Therefore, it is paramount to analyse the social media data to understand the human behavior and reaction in the ongoing pandemic.

To find out the useful insights from public opinions and shared posts in social media, and to model the public emotions, we have started collecting tweets from 5th March 2020. We have collected and processed over 200 million tweets related to Corona virus (focused on USA) which is about 1.3 terabytes data. We understand that processing this huge amount of data in real-time requires a substantial amount of time and resources. Hence, we decided to share the processed dataset with the research community so that they can skip the raw data processing steps and dive into data analytics and modeling. Thus, a cleaned dataset is readily available to the community and updated frequently. The dataset contains a set of useful features including processed tweet texts, user geo-location, and tweet IDs. In this work, we primarily,

\begin{itemize}
    \item Developed a web application in order to track,  collect and analyze tweets related to COVID-19.
    \item The web application contains several fundamental data analysis on the collected data such as topic trends, sentiment analysis, and user movement information. Those analysis is updated in real-time continuously and might be useful for the research community to understand the tweets better.

\end{itemize}

\section{Related Works}
Twitter has been proven useful for various tasks such as emergency communication network, public emotions monitor, detect anomalies and provide early warning, etc. Twitter was used as the data source to monitor the public reaction and health during disasters (e.g. hurricanes \cite{kabir2019deep, baer2012sandy, zou2019social, yang2019twitter, sebastian2019leveraging}, floods \cite{hirata2018flooding}, earthquakes \cite{earle2012twitter}, terrorist bombing \cite{buntain2016evaluating}, public health related misinformation propagation \cite{southwell2019misinformation, broniatowski2018weaponized, oyeyemi2014ebola} and others \cite{wang2019vulnerable, wladdimiro2016disaster, imran2016twitter}),  and disease outbreaks \cite{nagar2014case, szomszor2010swineflu, odlum2015can, dredze2016zika}. 
Researchers are trying to come forward with various ideas which involves the use of twitter data in various ways. Catherine et al. \cite{ordun2020exploratory} use twitter data to explore and illustrate five different methods to analyze the topics, key terms and features, information dissemination and propagation, and network behavior during COVID-19. The authors use pattern matching and topic modeling using Latent Dirichlet Allocation (LDA) in order to select twenty different topics on spreading of corona cases, healthcare workers, and personal protective equipment (PPE). Using various analysis the authors were able to detect high level topic trends, relation with the sudden news or briefing with the topic spikes and finally the authors tried to understand how the topics are evolving with the time. Alaa et al. \cite{abd2020top} also performed topic modeling using word frequencies and Latent Dirichlet Allocation (LDA) with the aim to identify the primary topics shared in the tweets related to the COVID-19.

Lisa et al. \cite{singh2020first} and Ramez et al. \cite{kouzy2020coronavirus} presented their works on misinformation propagation and quantification related to COVID-19 using twitter. The authors in \cite{kouzy2020coronavirus} conclude that there is an alarming rate of medical misinformation and non-credible content which are posted and shared in the twitter throughout the pandemic. It is very crucial to quantify the misinformation on the social media and take necessary action in order to prevent unnecessary anxiety and medically harmful methods to fight against COVID-19.

Many Researchers from different countries are trying to collect and share twitter datasets on COVID-19 \cite{781w-ef42-20, banda2020large, chen2020tracking}. Those shared datasets contain tweets on a variety of keywords from different countries. While some data repository is sharing only tweet IDs, others also sharing some form of prepossessed data. In this work, we also aim so sharing our collected dataset with the respected community along with tweet IDs and a cleaned version of processed tweets \cite{kabir2020coronavis} with some useful tweet attributes. The primary aim is to reduce the computational time and power of the researcher by providing a dataset that is readily available for further analysis. 

\section{Data Description}
\begin{figure*}
   \centering
    \includegraphics[width=7in]{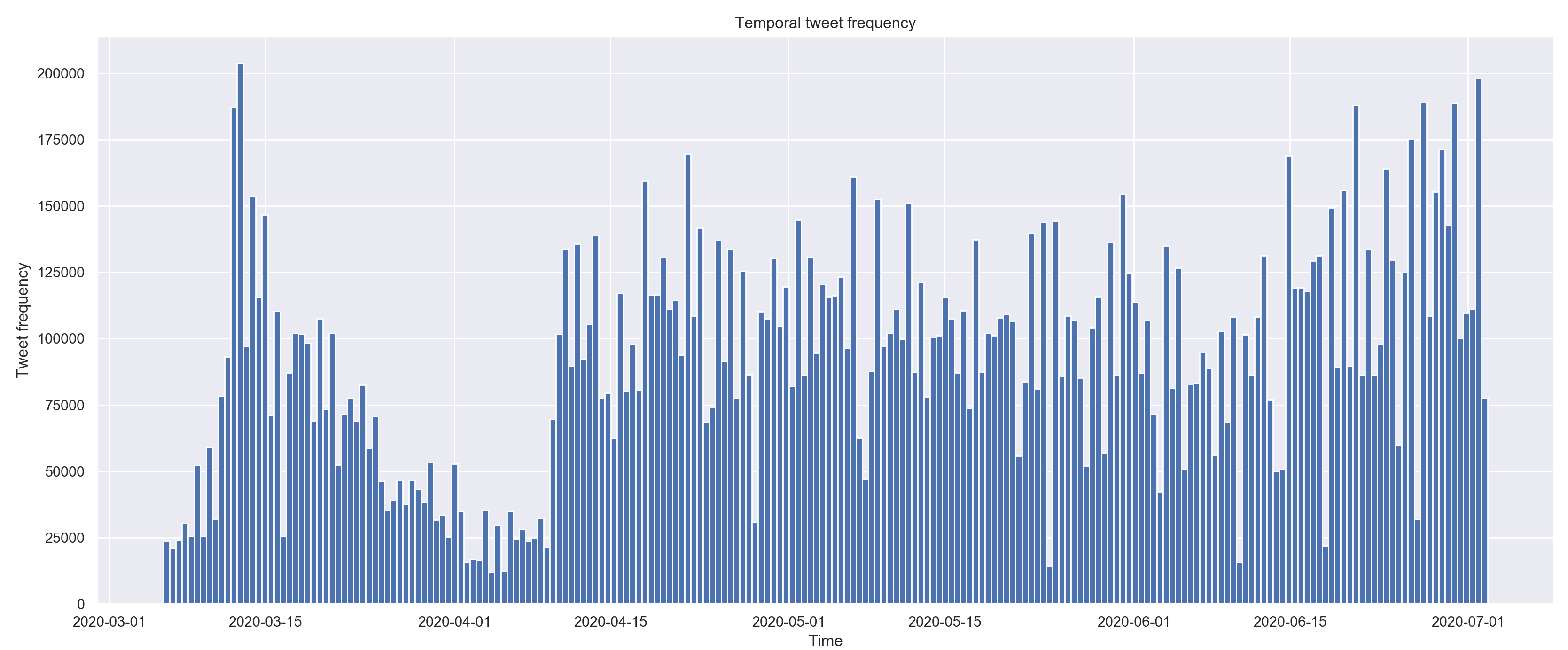}
    \caption{tweet frequency}
    \label{fig:tweet_frequency}
\end{figure*}

\subsection{Data Collection}
We are continuously collecting the data since March 5, 2020 and will keep fetching the tweets using Twitter Streaming API\footnote{\url{https://developer.twitter.com/en/docs/tutorials/consuming-streaming-data}} and python Tweepy\footnote{\url{https://www.tweepy.org/}} package. We have collected more than 200M tweets which has around 1.3TB of raw data until July 2nd, 2020 and saved this data as JSON files. We are using COVID-19 related keywords such as covid, corona for the tweets collection. The module listen the stream of the tweets and try to check if a tweet contains covid, or corona in it. While checking the module convert all the text to lower case and try to find out sub-string within the text. By doing this the module identify a qualified tweet and save of any tweet that contains covid or corona as a single word or as a part of the word. We also dynamically process this data in real-time for the CoronaVis\footnote{\url{https://mykabir.github.io/coronavis/}} application. We processed several features from the tweets such as Tweet ID, created time, Tweet Text, User geo-location if available, User Type, Sentiment (polarity and subjectivity), etc. For the live application tweets with geo-location information is saved separately for further data analysis. We will keep collecting the data and update the data repository (\url{https://github.com/mykabir/COVID19}) once in every week. The repository contains the processed data that is used for the live application and further analysis. In the following subsection, we briefly describe the attributes of the dataset. 

\subsection{Data Description}
Table \ref{dataummary} represents a high level data summary that is available in the git repository. However, we are continuously collecting the data and thus the data statistics can be changed in the repository with future updates. In the repository, we have included processed data only with the geo-location information. However, we will also include the list of all tweets ID with or without geo-location information.

\begin{table}[h]
\caption{Data Summary}
\label{dataummary}
\centering
\begin{tabular}{|p{2.25cm}|p{5.75cm}|}
\hline
\bfseries Attribute & \bfseries Summary \\ 
\hline
 First tweet time & Thu Mar 05 20:37:08 2020.  \\
\hline
 Last tweet time & Thu Jul 02 18:25:50 2020.  \\
\hline
Number of unique tweets & 198,48,179. \\
\hline
Location & USA (State label). \\
\hline
Number of Unique users & Total: 30,70,047; \newline Verified: 45,015; \newline Non verified: 30,25,032; \\
\hline
\end{tabular}
\end{table}

The processed data is saved and updated in the git repository within the folder named as "data". The data folder contains several csv files. Every file contains tweets for the particular date that is specified as the name of that file. For example, 2020-03-05.csv contains the tweets that was fetch on 5th March, 2020. The name was formated as Year-Month-Date. All the data file contains 6 different attributes ( tweet\_id, created\_at, loc, text, user\_id, verified). The data contains tweets only with the location information as we focused our analysis only on the US data. To keep the privacy, in both our application and in shared data, we introduced some data annonymization. We included tweet ID with the data, hence the researcher can re-fetch the original tweet if that tweet is still publicly available. However, we understand that a user might want to remove a tweet or made that tweet publicly unavailable. In such a situation, to ensure the privacy of that user, we processed the tweet text and user name so that it can not be directly searched and linked to a particular user.  

\begin{table}[h]
\caption{Data attributes}
\label{hyperparams}
\centering
\begin{tabular}{|p{2.25cm}|p{5.75cm}|}
\hline
\bfseries Feature & \bfseries Description \\ 
\hline
 tweet\_id & Unique ID of a tweet.  \\
\hline
created\_at & Creation time of a tweet. \\
\hline
loc & State level user location. \\
\hline
text & Processed tweet text. All the text are in small letters, non-English characters and few stop words are removed. \\
\hline
user\_id & Pseudo user id. The exact user name is transformed to a anonymous id to preserve the privacy of the user.  \\
\hline
verified & Denotes whether the tweet post is verified or not (1 or 0). \\
\hline
\end{tabular}
\end{table}

Table \ref{hyperparams} represents the feature attributes in the shared data with a description. Figure \ref{fig:tweet_frequency} depicts the temporal tweet frequency over the time. There are some gaps in the collected datasets due to API and connectivity issues. Hence, in some of the date we might have fetched a fairly low amount of tweets compared to original number of the tweets on that day. However, we are working to fill up those gapes using other publicly available data repositories. We will also perform more data analysis and update the data repository with the more updated information and insights. 

\section{Data Analytics}
We have performed various data analysis using the fetched tweets. We will continue further data analysis in order to produce useful insights that might helpful to understand and fight against the COVID-19 pandemic situation. Figures 1 to 10 represent some example of data analytics on the processed dataset. A brief description of how we produced those graphs and figure are given below. 

\subsection{Frequent words}
Figure \ref{fig:wordcount_bar} represent the most frequent words and their frequency in the tweets. We can observe that cases, trump, deaths, etc. are the most frequent words that appears in the tweets along with covid related keywords.
\begin{figure}[h]
   \centering
    \includegraphics[width=3.5in]{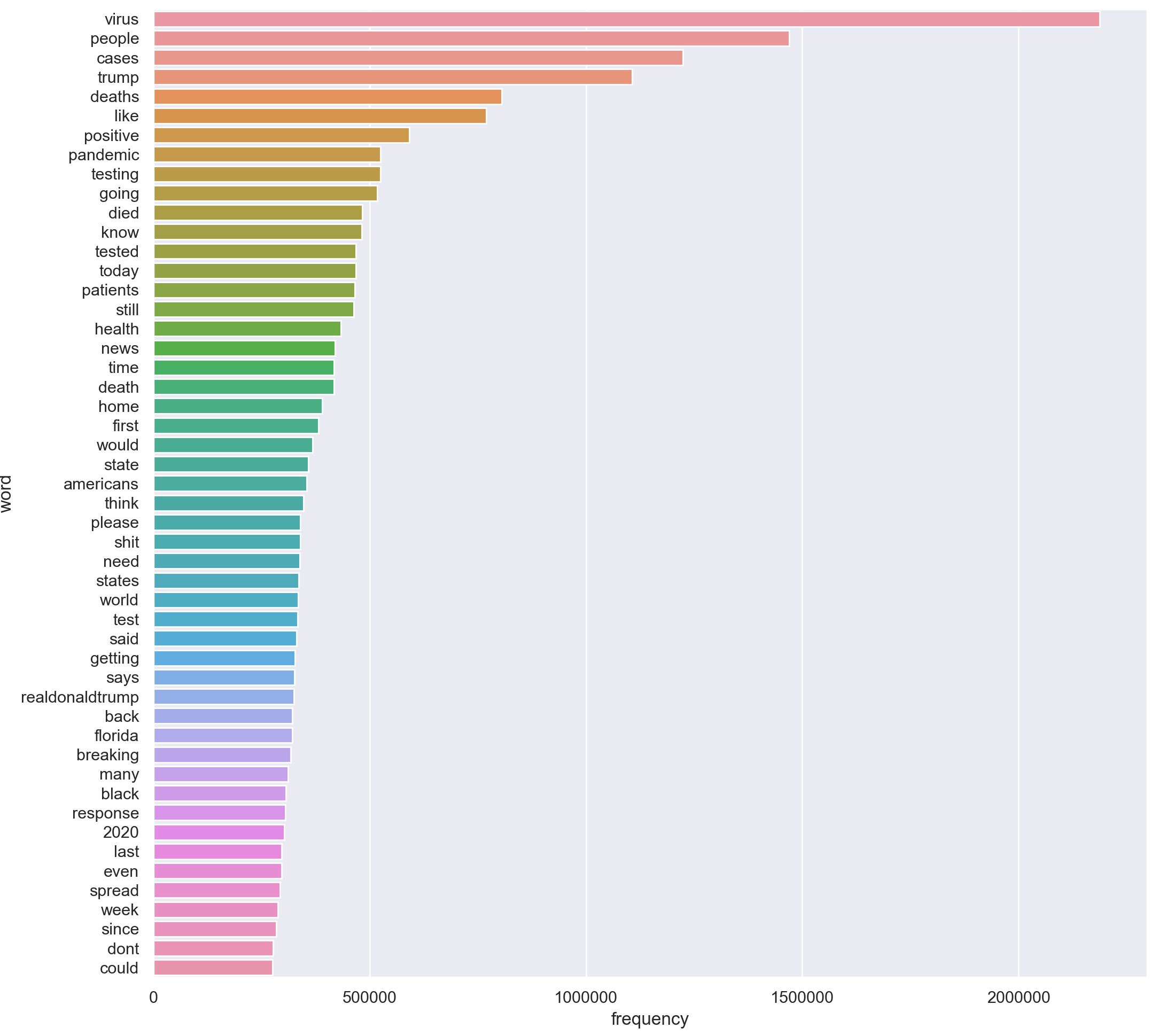}
    \caption{Word count}
    \label{fig:wordcount_bar}
\end{figure}

\begin{figure*}
   \centering
    \includegraphics[width=7in]{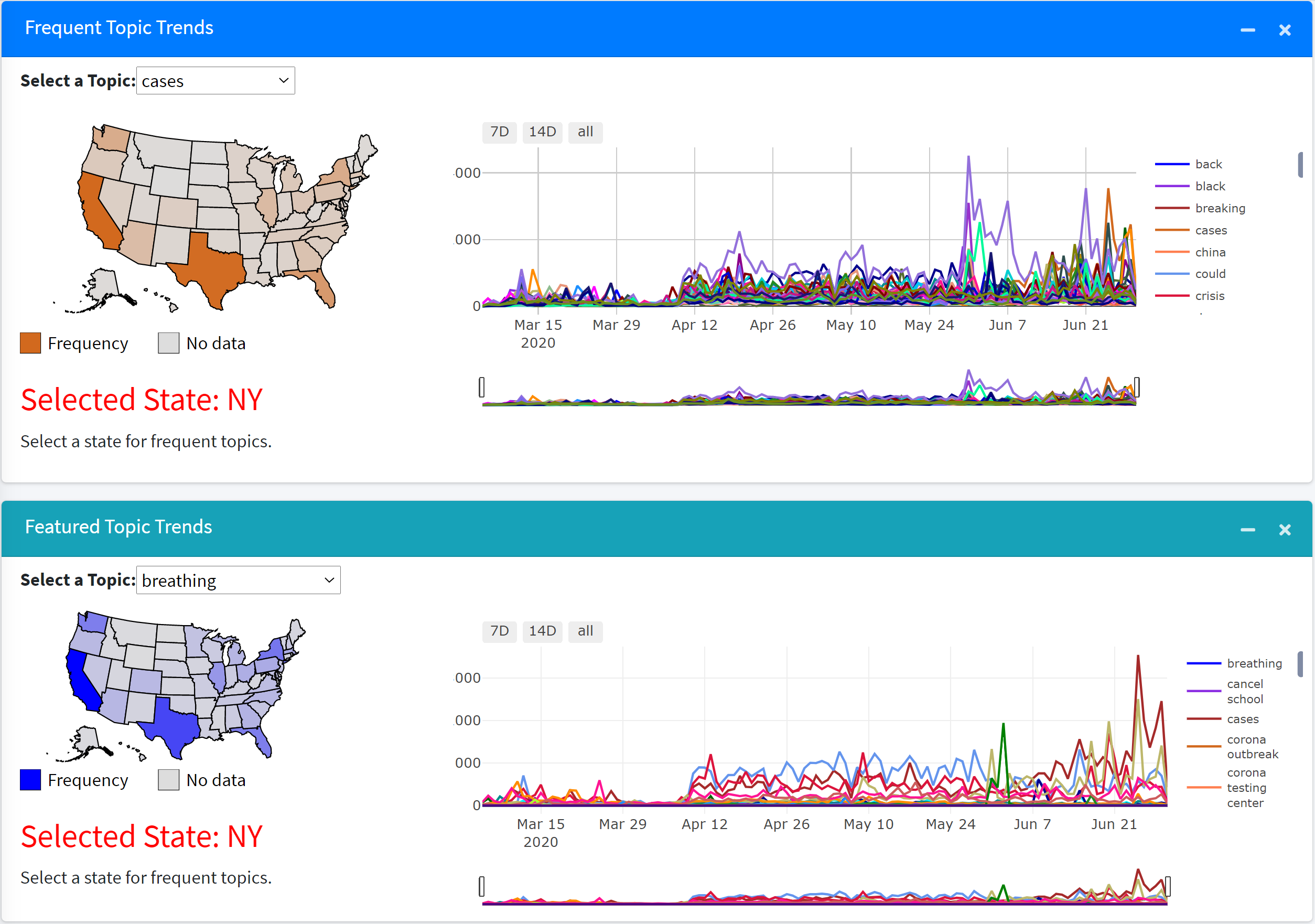}
    \caption{topic-trends}
    \label{fig:topictrends}
\end{figure*}

\subsection{Top Frequent Bigrams}
\begin{figure}[h]
   \centering
    \includegraphics[width=3.5in]{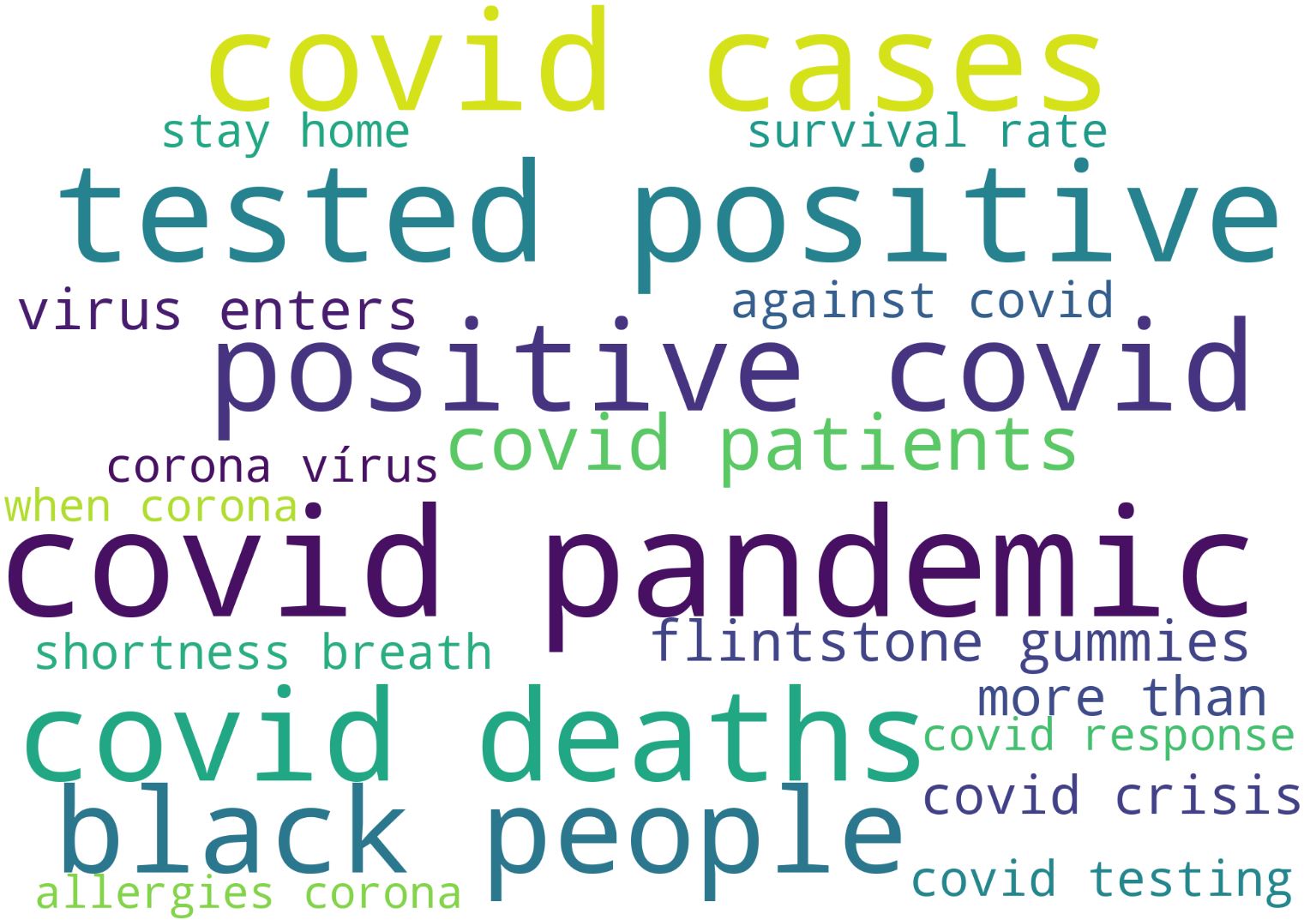}
    \caption{Top 20 frequent bigrams}
    \label{fig:bigrams}
\end{figure}

Figure \ref{fig:bigrams} represents a word cloud consists of the top 20 frequent bi-grams in the tweets. We consider all the tweets from March 5th to July 2nd. After removing the duplicates and noises (such as irrelevant symbols, stop words, urls, etc) we calculate the frequency of each pair of words in the all tweets text. The top frequent bigrams provides a brief idea of the topics that people are talking along with COVID related text. 

\subsection{Topic Trends}

Figure \ref{fig:topictrends} shows the most frequent topics, and the selected featured topics for New York (NY) state. From the figure, we can observe that the topic \textbf{cases} became highly frequent by the end of June which denotes that the infections is increasing again in all over the country, and we can also observe that the topics such as \textbf{mask, death} also became popular on the same time. We observe that the topic trends became different for different states in  USA. To observe how the topic trends are evolving, and what challenges the community are discussing, we have created two different graphs. First, Frequent topic trends and Second, Featured or selected topic trends. Our developed application continuously count the frequency of the words in the tweets and find the top 50 most frequent words till date. However, the application removes some of the stop words, and some other selected words that might be frequent but not meaningful (e.g. to, for, what, how, covid, corona, etc.) 

\begin{figure*}
   \centering
    \includegraphics[width=7in]{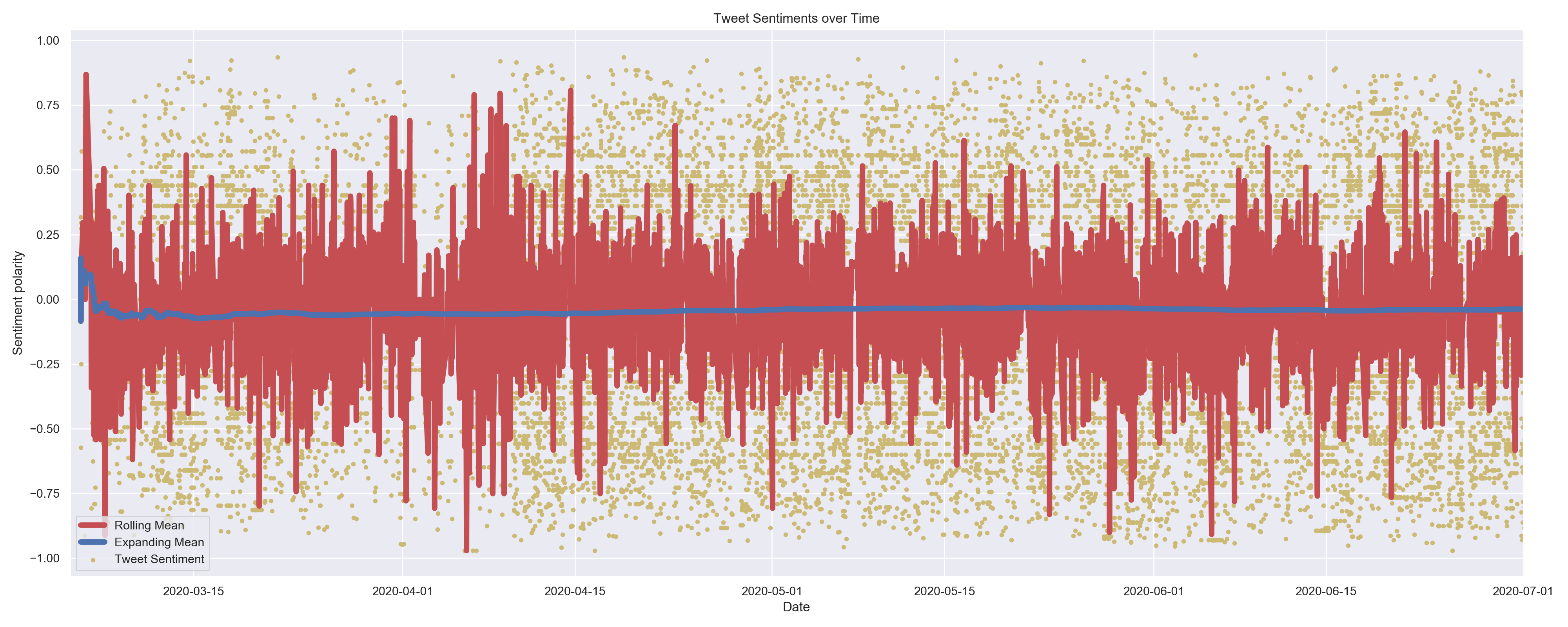}
    \caption{Temporal sentiments from March 5th to July 1st, 2020}
    \label{fig:temporal_sentiments}
\end{figure*}

Further, using python plotly package \cite{sievert2017plotly} the application creates the interactive charts which allows a user to observe single, multiple, or all topic trends together. For featured topic trends, we manually created an array containing 100 topics of our interest (e.g. breathing, cases, testing, corona outbreak, need help, etc.). The application keep tracks of those topics and rank them based on the frequency of the words and finally display the top 50 topics in the line chart. The module can produce frequent and featured topic trends charts for the whole USA and individual states. 

\subsection{Sentiment (Polarity)}
With the aim to understand the people reaction during the COVID-19 pandemic, we performed extensive analysis on the sentiment of the shared tweets and the users. We perfomed all the analysis using 3 categories (All tweets, tweets by verified profile, and tweets by non verified profile). We found that there is good contrast between the content shared by the verified users vs non verified users. Figure \ref{fig:temporal_sentiments} represents the temporal sentiments in the collected tweets. We can see that the polarity of the sentiments is distributed across the scale while it is mostly ranged between -.40 to .40. However, in some of days we can see a spike in positive of negative polarity.

\begin{figure}[h]
   \centering
    \includegraphics[width=3.5in]{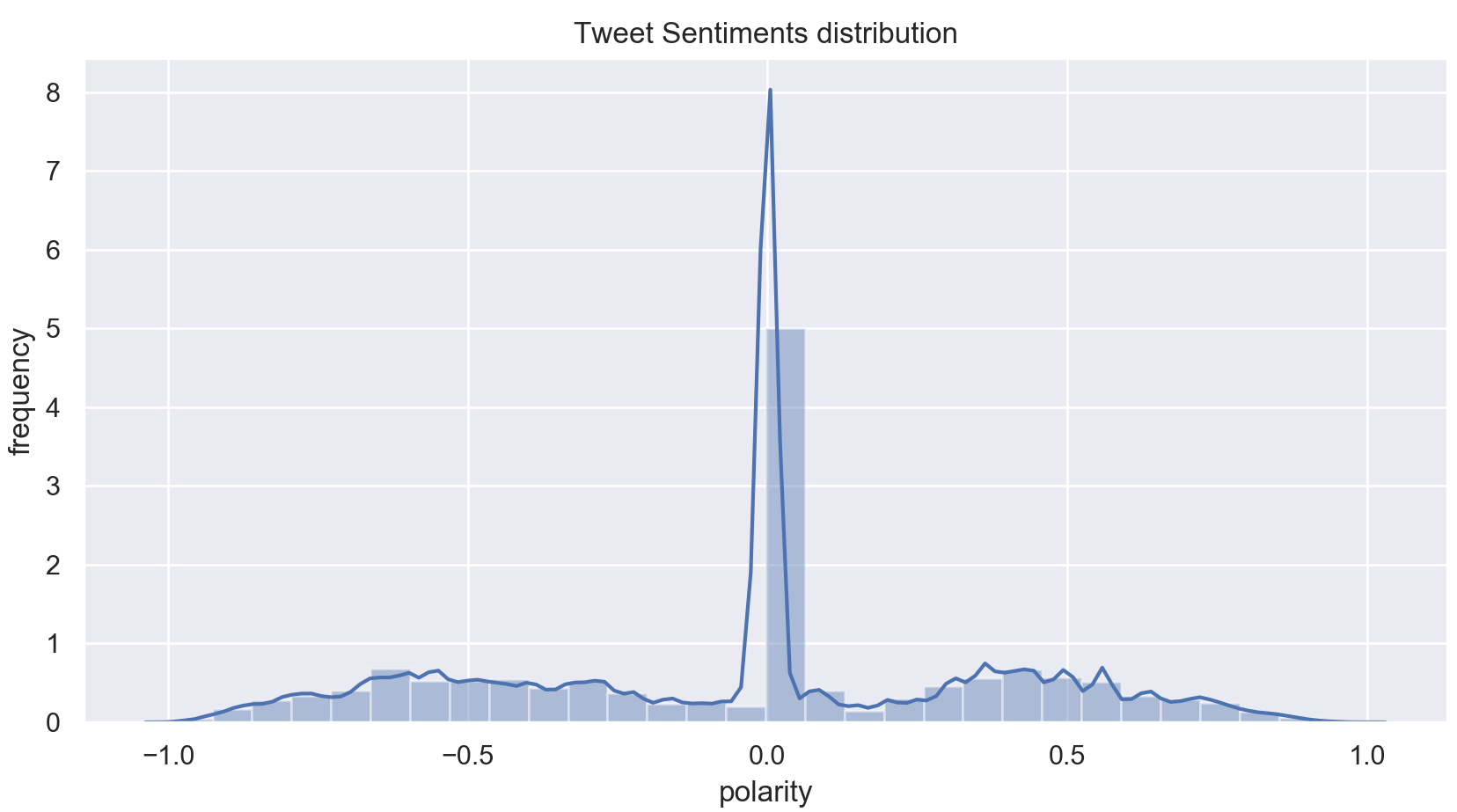}
    \caption{sentiment distribution}
    \label{fig:sentiment_distribution}
\end{figure}

Figure \ref{fig:sentiment_distribution} represents the sentiment (polarity) distributions in all the tweets. We can see that most of the tweets are in neutral zone with a bit of positive polarity. However, it is clear that there is more aggressive sentiments in the tweets around -0.4 to -.6 compared to the positive counter part. 

\begin{figure}[h]
   \centering
    \includegraphics[width=3.4in]{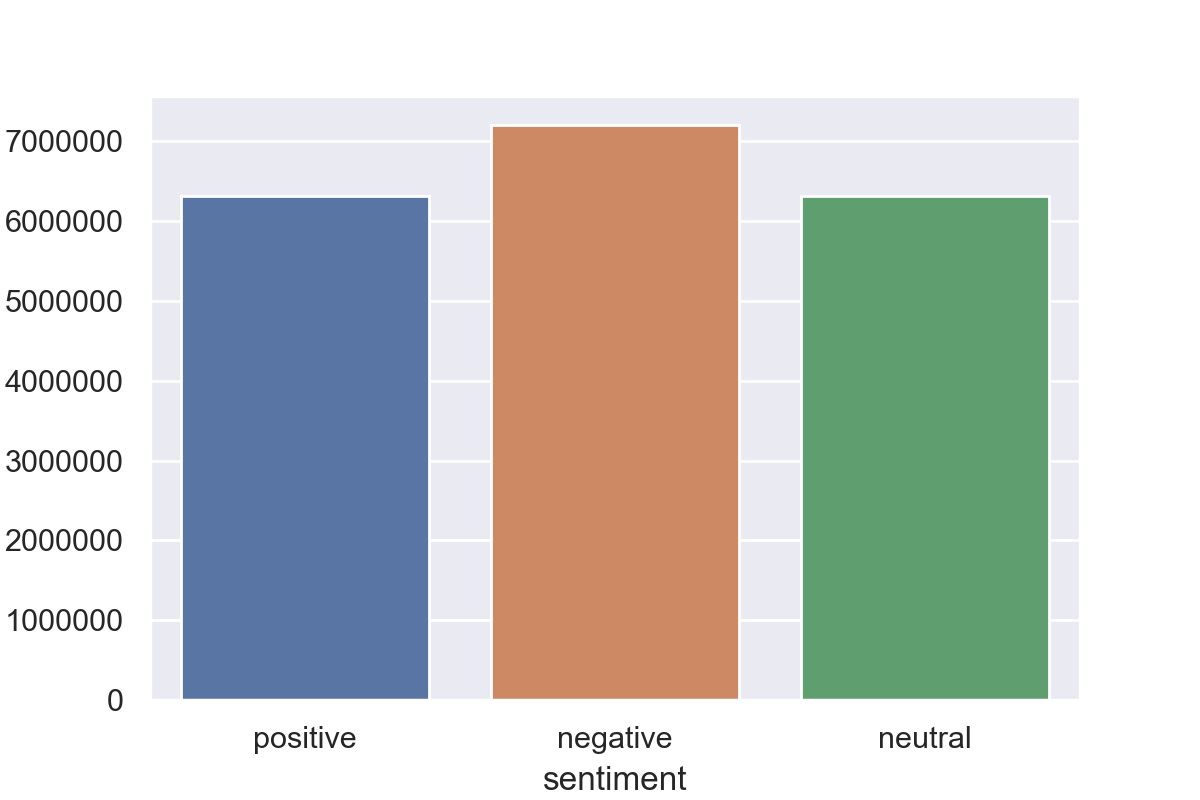}
    \caption{Sentiments distributions}
    \label{fig:sentiment}
\end{figure}

The number of positive, negative and neutral tweets are presented in figure \ref{fig:sentiment}. We use the Sentiment intensity analyzer from python Natural Language Toolkit (NLTK) library package \cite{loper2002nltk}. We use the compound score which is a normalized score. A sentiment between -0.05 to +0.05 is considered as neural and others as either positive or negative. We can see that the number of tweets with negative polarity is fairly larger in the data. 

\begin{figure}[h]
   \centering
    \includegraphics[width=3.5in]{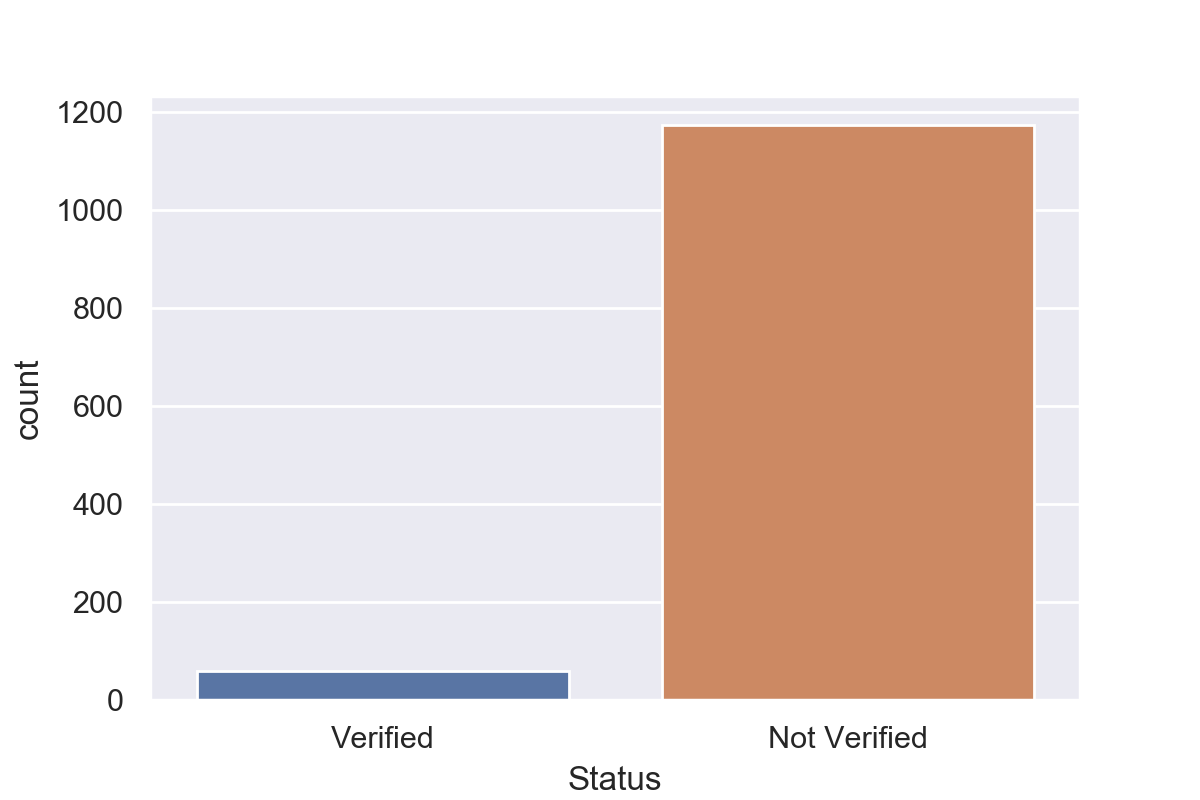}
    \caption{Verified vs non-verified users with more than 500 tweets}
    \label{fig:v_vs_nv}
\end{figure}

To visualize and understand the contrast between verified and non-verified users tweets we purposefully selected the tweets from user who tweeted more than 500 times during this period. We wanted to observe what kind of sentiment the frequent users are presenting in the social network. Figure \ref{fig:v_vs_nv} shows the number of verified and non-verified users with more than 500 tweets on COVID-19. The userId can be a person or entity. 

\begin{figure}[h]
   \centering
    \includegraphics[width=3.5in]{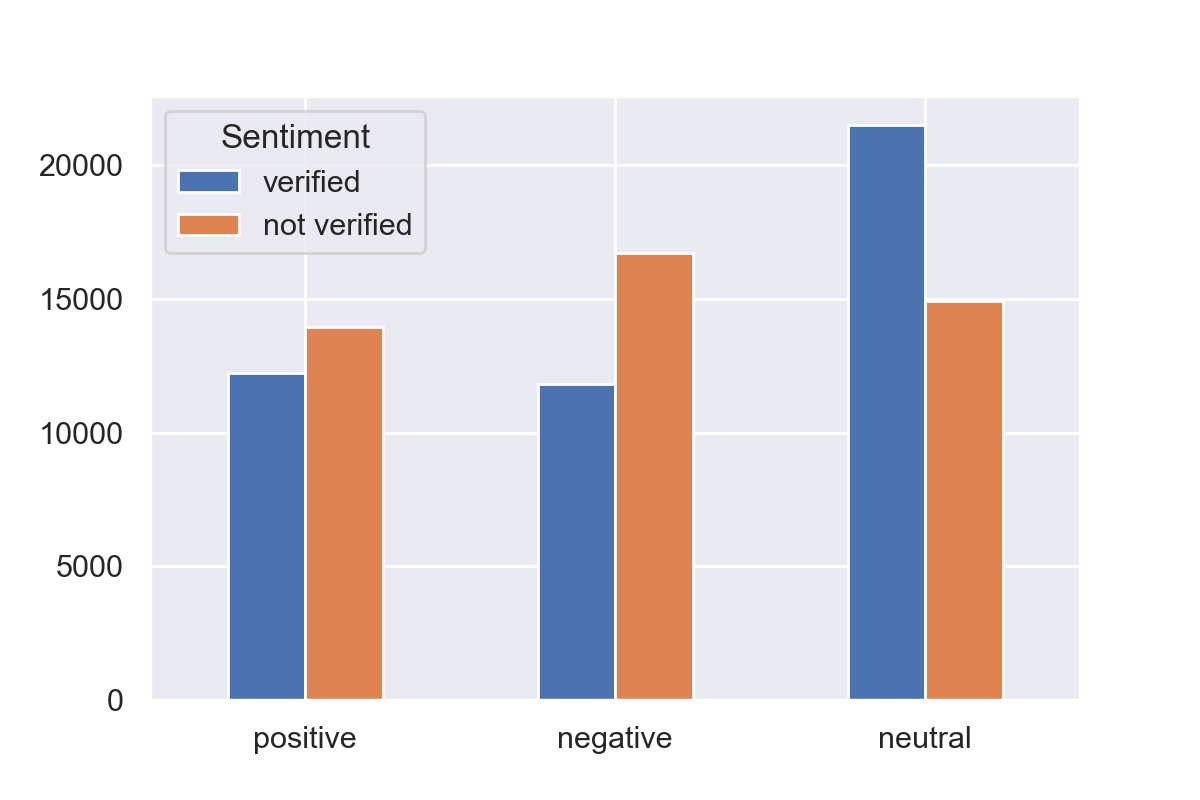}
    \caption{Sentiments distributions verified vs non-verified}
    \label{fig:sentiment_v_nv}
\end{figure}

The sentiments distribution between verified vs not-verified profile users is depicted in figure \ref{fig:sentiment_v_nv}. There are 45,547 tweets from verified profile (only 61) and 9,26,963 tweets from the users who is not verified (1174). The maximum number of tweets from a single verified profile is 1873 where 8493 tweets were posted by a single profile which is not verified. To represents the number of tweets with positive, negative and neutral polarity we normalize the number of tweets in figure \ref{fig:sentiment_v_nv}. It is clearly observable that profiles without verification are posting tweets with higher negative polarity. The verified profiles tends to share content with a neutral tone. 

\begin{figure}[h]
   \centering
    \includegraphics[width=3.4in]{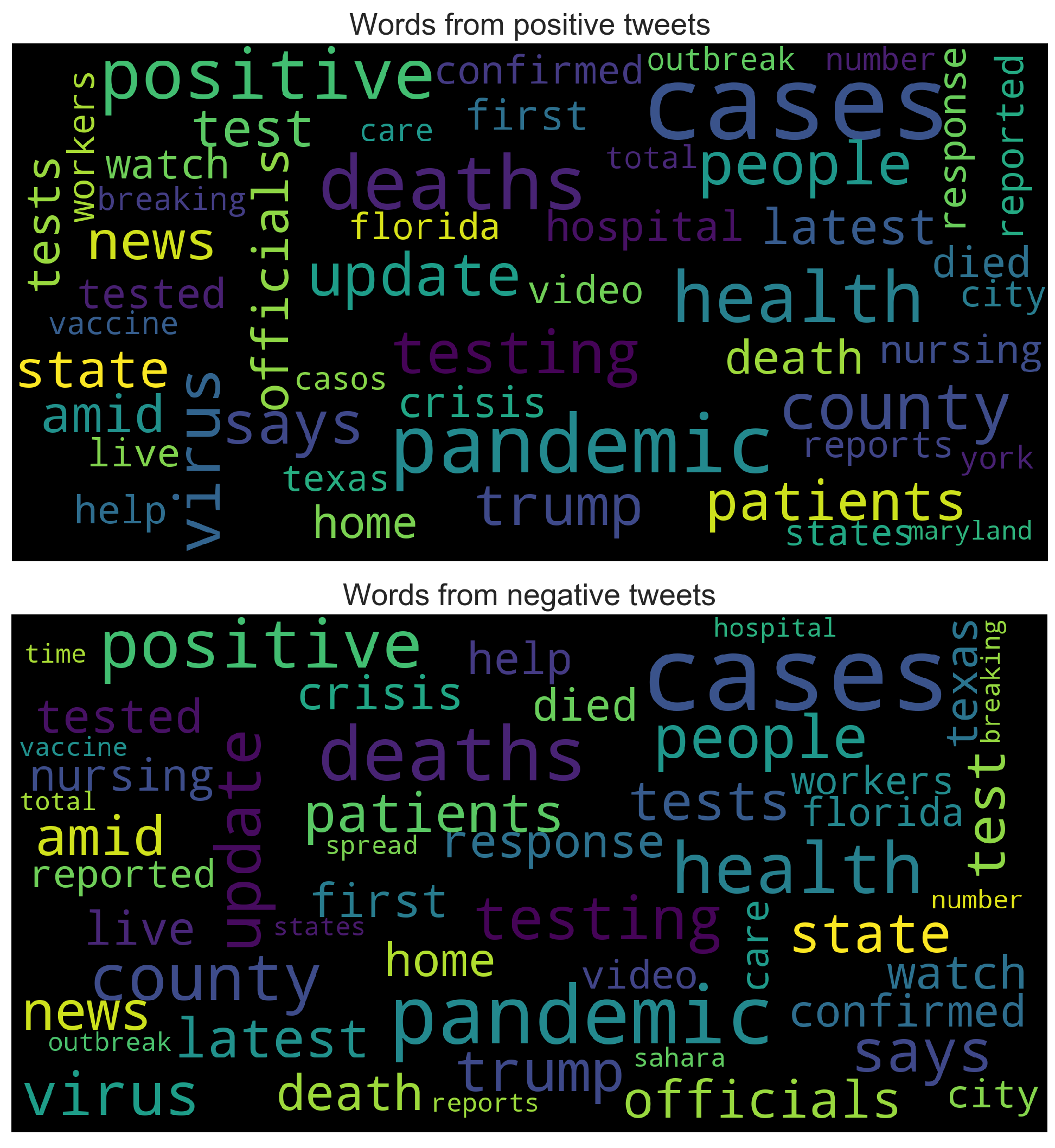}
    \caption{Positive and Negative word cloud from verified tweets}
    \label{fig:wc_v}
    \squeezeup
\end{figure}

\begin{figure}[h]
   \centering
    \includegraphics[width=3.4in]{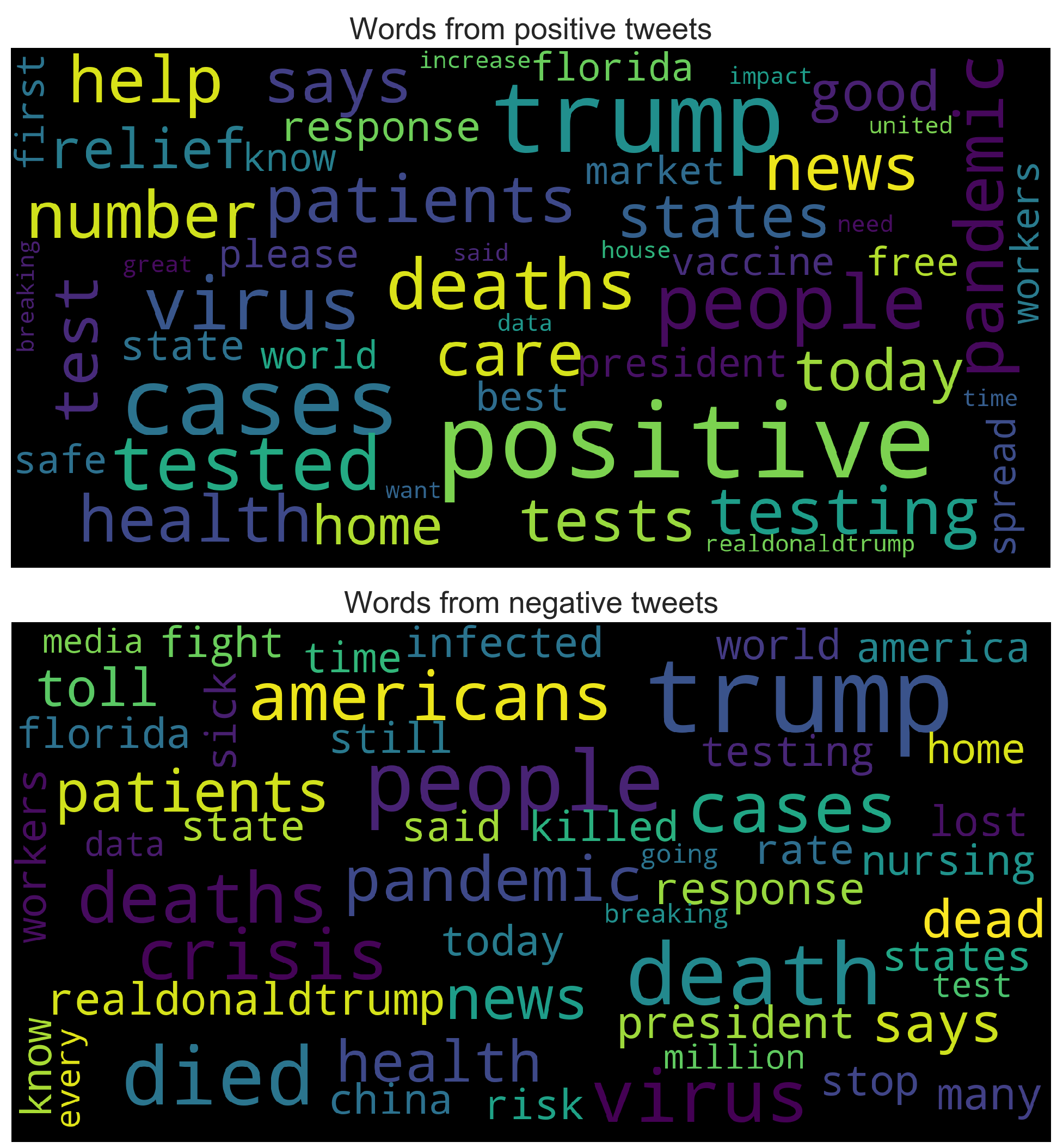}
    \caption{Positive and Negative word cloud from non-verified tweets}
    \label{fig:wc_nv}
\end{figure}

Positive and negative word-clouds are presented in figure \ref{fig:wc_v} and figure \ref{fig:wc_nv} using the words from the tweets by verified and non-verified profiles. The word-clouds contain the most frequent 50 words from each category of the tweets. We can observe that most of the words are overlapped across the figures. However, there is some distinct differences among the word-clouds. For example, "trump, deaths/die/died " appears more frequently in non-verified tweets. On the other hand "update, officials" are more frequent in the verified tweet texts. 

\begin{figure}[h]
   \centering
    \includegraphics[width=3.5in]{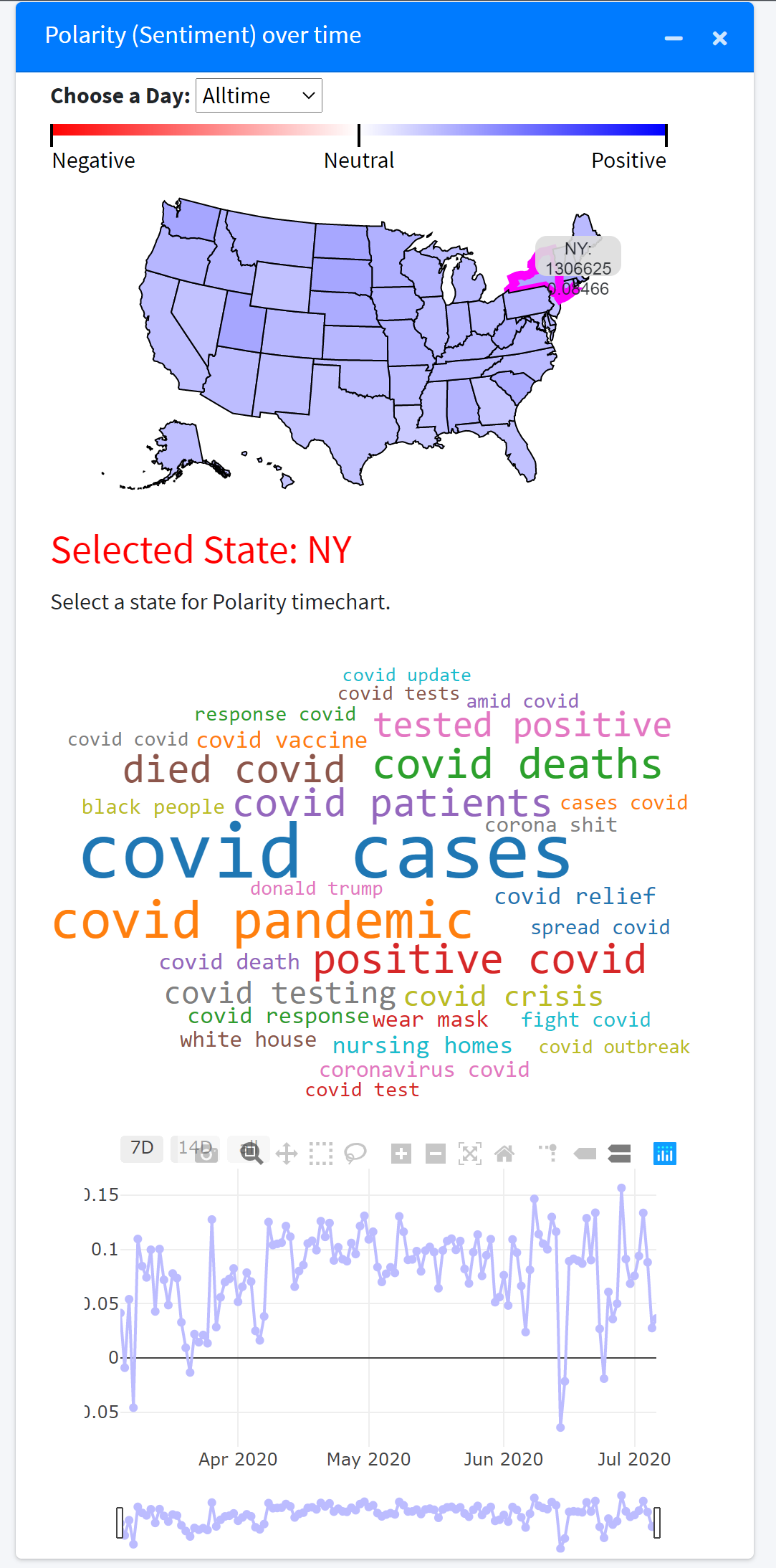}
    \caption{Polarity over time in New York (NY) }
    \label{fig:polarityny}
\end{figure}

\begin{figure}[h]
   \centering
    \includegraphics[width=3.5in]{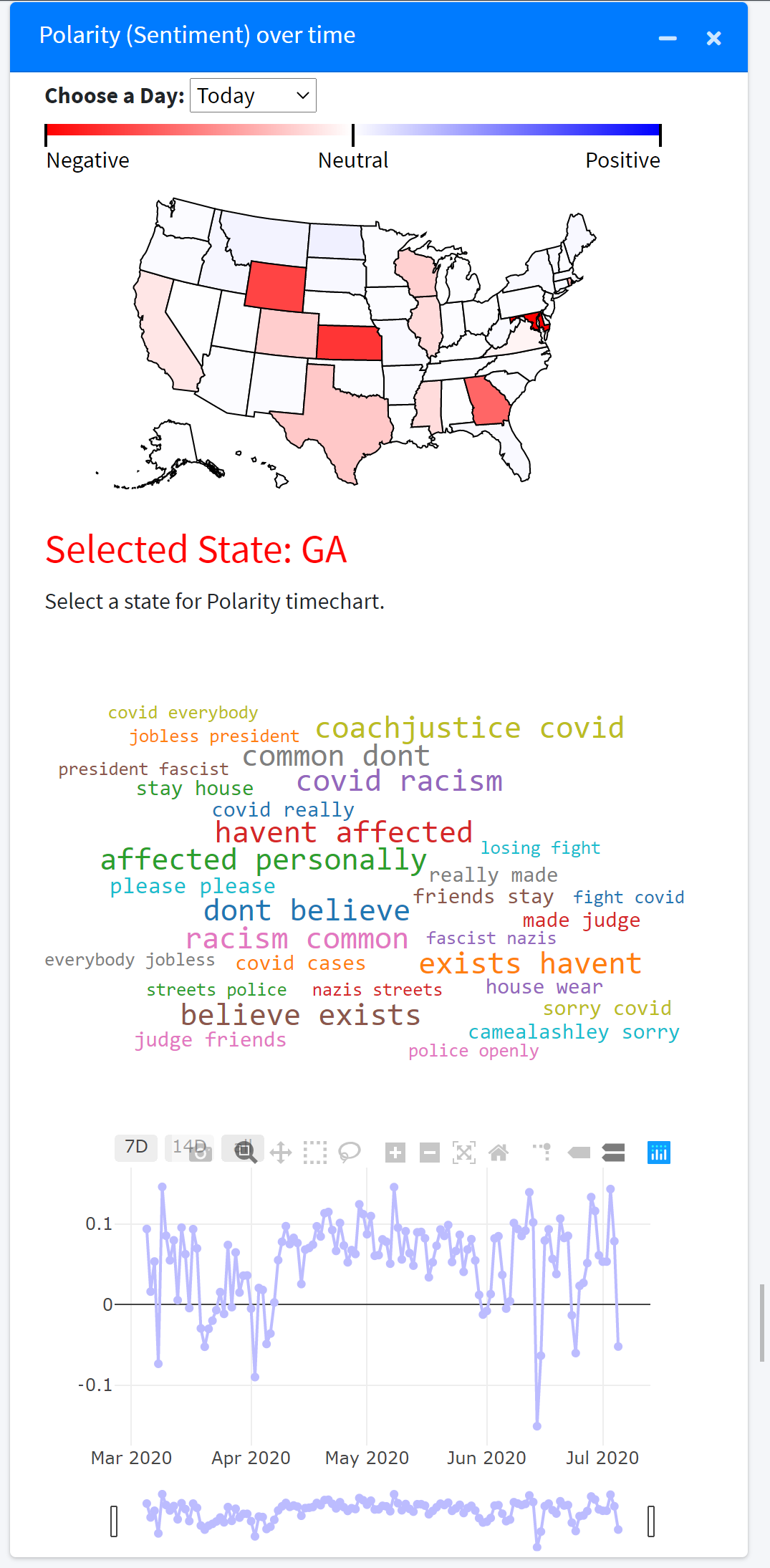}
    \caption{Polarity over time in Georgia (GA)}
    \label{fig:polarityga}
\end{figure}

Polarity over time for different states is depicted in figure \ref{fig:polarityny}. In the figure the top image represents a clickable geo-map where a user can select a state and observe polarity (sentiment) of that state over a specific time period. The user can select a time frame from the drop down menu (e.g. Today, Yesterday. All time). The application process and load the respective charts based on the selection. The word cloud (at the middle) provides an idea of what has driven the sentiments in that state for the selected time. The word cloud represents the most frequent bigrams in the tweets for that particular states. The selected time frame also reflected in word cloud and time chart. The bottom image of the figure shows the daily polarity over time. In figure \ref{fig:polarityny}, the polarity analysis for New York (NY) states is presented. In Figure \ref{fig:polarityga} we can observe the Today sentiment (5th July, 2020) for the state Georgia (GA). The geo-map is clearly showing a negative sentiment across the USA. 

\begin{figure*}
   \centering
    \includegraphics[width=7in]{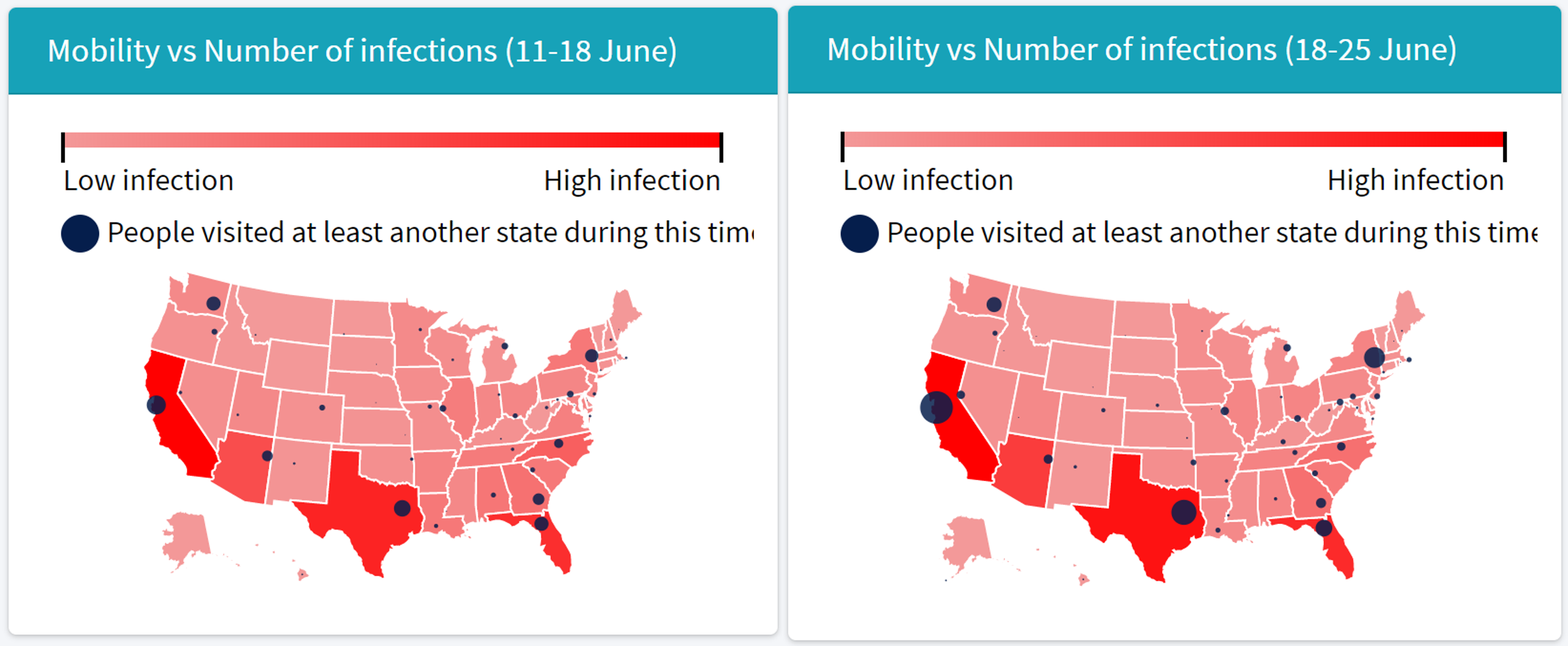}
    \caption{Weekly mobility vs Number of infections}
    \label{fig:mobility}
\end{figure*}

\begin{figure}[h]
   \centering
    \includegraphics[width=3.4in]{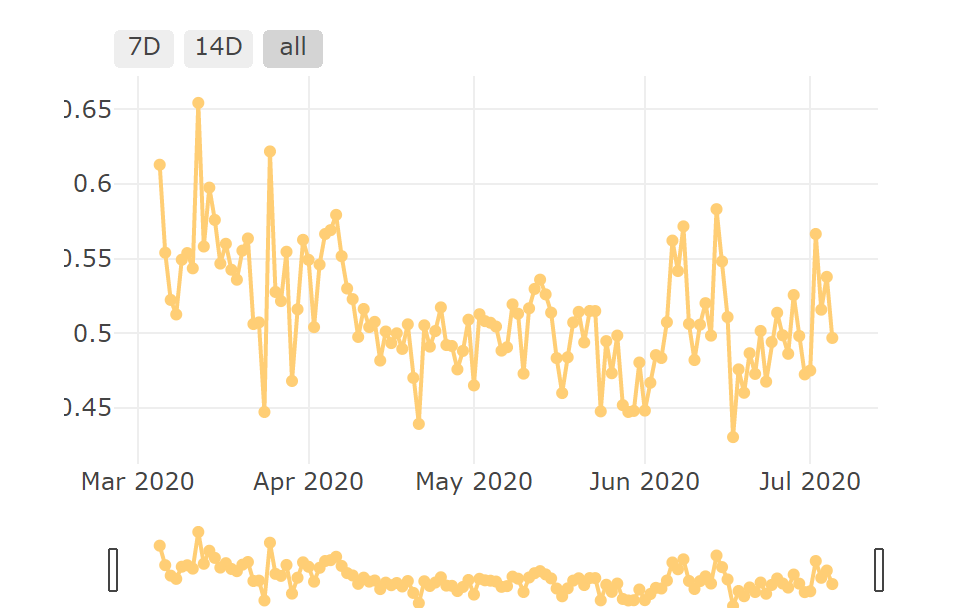}
    \caption{Subjectivity in the tweets from Georgia}
    \label{fig:subjectivity}
\end{figure}

The negative sentiment relation is well co-related with the present spikes of new cases in different states. For instance, GA has more than 2500 new patients on this date. From the word cloud we can observe an ongoing frustration among the people. The CoronaVis application also allows observing the subjectivity analysis of any state in a similar fashion. Subjectivity is a measurement of fact or opinion in a text ranging from 0 to 1.   Figure \ref{fig:subjectivity} shows the subjectivity of COVID-19 related tweet for the state Georgia. We can observe that the aggregate texts of the tweets shows a subjectivity in the middle of fact and opinion. However, for some of the date people share more factual information.

\subsection{User Movement}
COVID-19 virus is highly contagious and it spread fast with the movement of the people. In this module we analyze the correlation between people mobility and number of infection using the twitter data. We observe the number of cases and user movement between multiple states in every week. As our data contains high level geo-graphical information we can only observe the state label user mobility. If a user moves between two or more states within 14 days we consider that as mobility count. For example if a person post a tweet from NY and within next 14 days if that person visit another state such as Pennsylvania (PA) we count that person for the mobility calculation for PA state. 

\begin{figure}[h]
   \centering
    \includegraphics[width=3.5in]{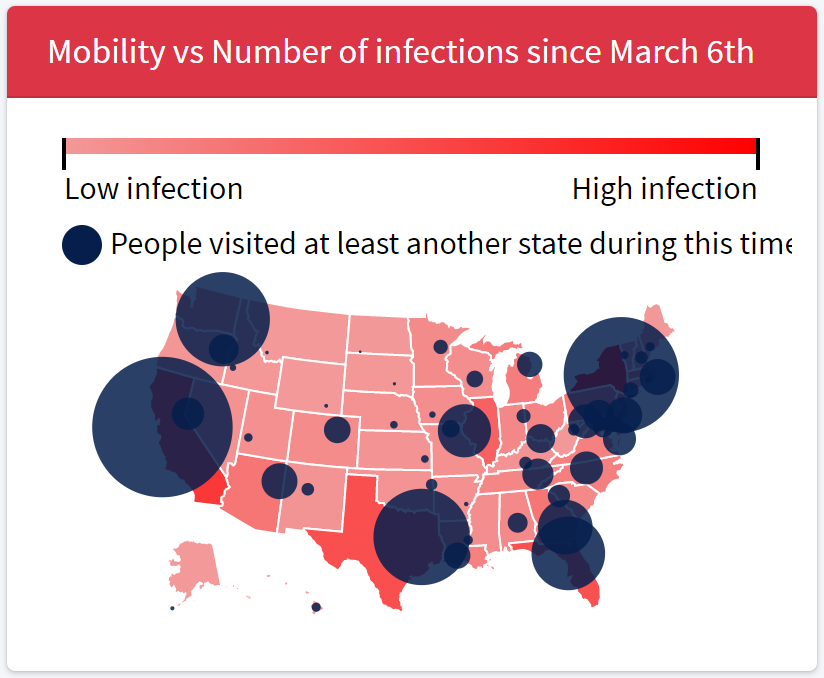}
    \caption{Mobility vs Number of infections}
    \label{fig:usermap}
\end{figure}

The number of the users moving between two or more states during 11th June to 18th June and 18th June to 25th June along with the number of corona virus cases in those states are presented in figure \ref{fig:mobility}. As we know that a person might take 1 week or more before showing symptoms after getting infection the mobility count in the figure showing the mobility of the previous week. For example, Mobility vs Number of infections (18-25 June) represents the user mobility between 11-18th June while presenting the number of infection for the week of 18-25th June. From the figure we can observe that mobility has a good impact on the COVID-19 propagation. 

\begin{figure*}
   \centering
    \includegraphics[width=7in]{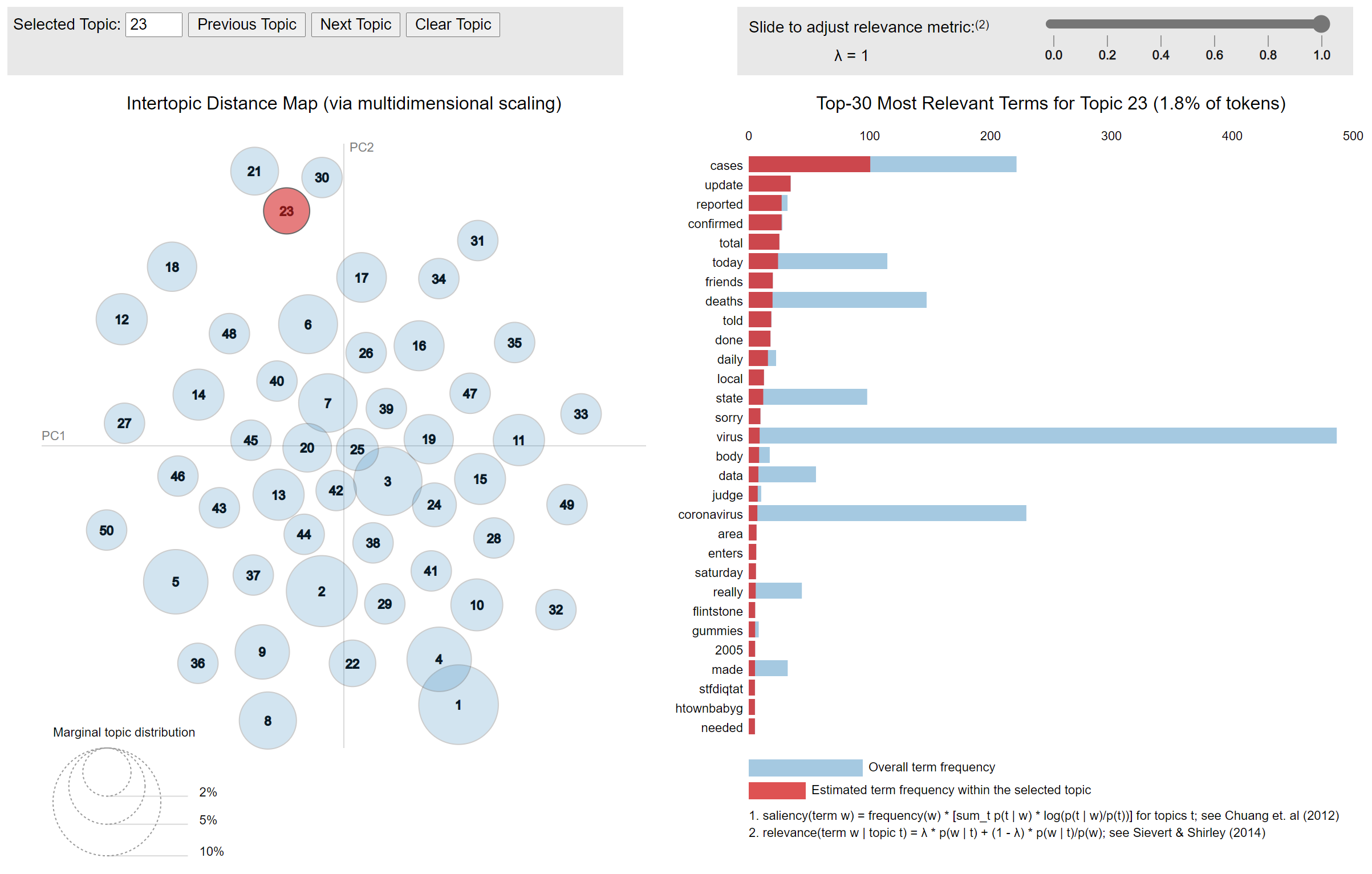}
    \caption{Topic Modeling: most relevant terms for topic "cases"}
    \label{fig:lda}
\end{figure*}

We can see that places (e.g. Texas, Florida) with increased mobility are experiencing greater number of infection in last few weeks. Figure \ref{fig:usermap} depicts the mobility vs number of infection since March 6th, 2020. The live application allows a user to hover on the bubbles or map to see the exact number of the infections and moving people. Although there are a number of factors that contribute to the total number of infections in a location, however the higher mobility evidently correlates well with the infections. 

\subsection{Topic Modeling}
Topic modeling can be useful to tell what are the relevant topics that is appearing in the tweets along with the top topics. We use Latent Dirichlet Allocation (LDA) and python pyLDAvis package to find out the relevant topics and produce an interactive visualization. For the visualization we remove the common stop words from the tweet corpus along with some specific words such as covid, corona, RT, etc. Then the corpus is tokenized and transformed using TF-IDF model \cite{ramos2003using}. Further using LDA another transformation is performed from bag-of-words counts into a lower dimensionality topic space. Figure \ref{fig:lda} shows the top-30 most relevant terms for Topic 23 which is "cases". An interactive version of the topic visualization is accessible at \textbf{\url{https://mykabir.github.io/coronavis/lda.html}}

\section{Data Access and Use Policy}
\subsection{Data Access}
The data is accessible from the Github repository \textbf{\url{https://github.com/mykabir/COVID19}}. The repository contains two different folders. One is the \textbf{data} folder containing the tweeter data and another is a \textbf{src} folder which contains some basic code presenting the way to read the data and some basic data analytics. The \textbf{data} folder contains the data in CSV file format for each day from 5th March 2020 to till date and named by the particular date with the format YEAR-MONTH-DATE. The live application (CoronaVis) is accessible at \textbf{\url{https://mykabir.github.io/coronavis/}}. We are working continuously to add more real-time data visualization and analytics on the website. If you have any suggestions or concern, please send an email at mkabir@mst.edu or madrias@mst.edu.

\subsection{Use Policy}
This dataset is released in compliance with Twitter’s Developer
Terms \& Conditions\footnote{\url{https://developer.twitter.com/en/developer-terms/agreement-and-policy}}. The data repository will be continuously updated every week. The data repository, containing codes, and CoronaVis\footnote{\url{https://mykabir.github.io/coronavis}}, 2020 W2C lab, Missouri University of Science and Technology, all rights reserved, can be used for educational, academic, and government research purposes with proper citation (Please cite this paper). Any commercial use of any materials is strictly prohibited. Taking and sharing a screenshot is allowed with appropriate citation.  

\section{Conclusion}
In this work, we have developed a interactive web application for real-time tweets tracking on COVID-19 and producing insights dynamically. We performed sentiment analysis and related that with trending topics to find out the reason behind a sentiment for better understanding of the human emotions. We shared our processed dataset publicly so that the research community can take advantage of the data and keep contribution to fight against COVID-19. Currently, we our working to develop machine learning models to detect and understand the dominant emotion of the people using tweets related to COVID-19. We are labelling data manually and we will shared that dataset also with the community for further research.



\bibliographystyle{./bibliography/IEEEtran}
\bibliography{./bibliography/IEEEabrv,./bibliography/IEEEexample}

\begin{thebibliography}{10}
\providecommand{\url}[1]{#1}
\csname url@samestyle\endcsname
\providecommand{\newblock}{\relax}
\providecommand{\bibinfo}[2]{#2}
\providecommand{\BIBentrySTDinterwordspacing}{\spaceskip=0pt\relax}
\providecommand{\BIBentryALTinterwordstretchfactor}{4}
\providecommand{\BIBentryALTinterwordspacing}{\spaceskip=\fontdimen2\font plus
\BIBentryALTinterwordstretchfactor\fontdimen3\font minus
  \fontdimen4\font\relax}
\providecommand{\BIBforeignlanguage}[2]{{%
\expandafter\ifx\csname l@#1\endcsname\relax
\typeout{** WARNING: IEEEtran.bst: No hyphenation pattern has been}%
\typeout{** loaded for the language `#1'. Using the pattern for}%
\typeout{** the default language instead.}%
\else
\language=\csname l@#1\endcsname
\fi
#2}}
\providecommand{\BIBdecl}{\relax}
\BIBdecl

\bibitem{chen2015crime}
X.~Chen, Y.~Cho, and S.~Y. Jang, ``Crime prediction using twitter sentiment and
  weather,'' in \emph{2015 Systems and Information Engineering Design
  Symposium}.\hskip 1em plus 0.5em minus 0.4em\relax IEEE, 2015, pp. 63--68.

\bibitem{gerber2014predicting}
M.~S. Gerber, ``Predicting crime using twitter and kernel density estimation,''
  \emph{Decision Support Systems}, vol.~61, pp. 115--125, 2014.

\bibitem{grover2019polarization}
P.~Grover, A.~K. Kar, Y.~K. Dwivedi, and M.~Janssen, ``Polarization and
  acculturation in us election 2016 outcomes--can twitter analytics predict
  changes in voting preferences,'' \emph{Technological Forecasting and Social
  Change}, vol. 145, pp. 438--460, 2019.

\bibitem{kabir2019deep}
M.~Y. Kabir and S.~Madria, ``A deep learning approach for tweet classification
  and rescue scheduling for effective disaster management,'' in
  \emph{Proceedings of the 27th ACM SIGSPATIAL International Conference on
  Advances in Geographic Information Systems}, 2019, pp. 269--278.

\bibitem{baer2012sandy}
D.~Baer, ``As sandy became\# sandy, emergency services got social,'' \emph{Fast
  Company}, vol.~9, 2012.

\bibitem{zou2019social}
L.~Zou, N.~S. Lam, S.~Shams, H.~Cai, M.~A. Meyer, S.~Yang, K.~Lee, S.-J. Park,
  and M.~A. Reams, ``Social and geographical disparities in twitter use during
  hurricane harvey,'' \emph{International Journal of Digital Earth}, vol.~12,
  no.~11, pp. 1300--1318, 2019.

\bibitem{yang2019twitter}
J.~Yang, M.~Yu, H.~Qin, M.~Lu, and C.~Yang, ``A twitter data credibility
  framework—hurricane harvey as a use case,'' \emph{ISPRS International
  Journal of Geo-Information}, vol.~8, no.~3, p. 111, 2019.

\bibitem{sebastian2019leveraging}
A.~Sebastian, W.~HighField, S.~Brody, and W.~Mobley, ``Leveraging machine
  learning and twitter data to identify high hazard areas during hurricane
  harvey,'' 2019.

\bibitem{hirata2018flooding}
E.~Hirata, M.~Giannotti, A.~Larocca, and J.~Quintanilha, ``Flooding and
  inundation collaborative mapping--use of the crowdmap/ushahidi platform in
  the city of sao paulo, brazil,'' \emph{Journal of Flood Risk Management},
  vol.~11, pp. S98--S109, 2018.

\bibitem{earle2012twitter}
P.~S. Earle, D.~C. Bowden, and M.~Guy, ``Twitter earthquake detection:
  earthquake monitoring in a social world,'' \emph{Annals of Geophysics},
  vol.~54, no.~6, 2012.

\bibitem{buntain2016evaluating}
C.~Buntain, J.~Golbeck, B.~Liu, and G.~LaFree, ``Evaluating public response to
  the boston marathon bombing and other acts of terrorism through twitter,'' in
  \emph{Tenth International AAAI Conference on Web and Social Media}, 2016.

\bibitem{southwell2019misinformation}
B.~G. Southwell, J.~Niederdeppe, J.~N. Cappella, A.~Gaysynsky, D.~E. Kelley,
  A.~Oh, E.~B. Peterson, and W.-Y.~S. Chou, ``Misinformation as a misunderstood
  challenge to public health,'' \emph{American journal of preventive medicine},
  vol.~57, no.~2, pp. 282--285, 2019.

\bibitem{broniatowski2018weaponized}
D.~A. Broniatowski, A.~M. Jamison, S.~Qi, L.~AlKulaib, T.~Chen, A.~Benton,
  S.~C. Quinn, and M.~Dredze, ``Weaponized health communication: Twitter bots
  and russian trolls amplify the vaccine debate,'' \emph{American journal of
  public health}, vol. 108, no.~10, pp. 1378--1384, 2018.

\bibitem{oyeyemi2014ebola}
S.~O. Oyeyemi, E.~Gabarron, and R.~Wynn, ``Ebola, twitter, and misinformation:
  a dangerous combination?'' \emph{Bmj}, vol. 349, p. g6178, 2014.

\bibitem{wang2019vulnerable}
Z.~Wang, N.~S. Lam, N.~Obradovich, and X.~Ye, ``Are vulnerable communities
  digitally left behind in social responses to natural disasters? an evidence
  from hurricane sandy with twitter data,'' \emph{Applied geography}, vol. 108,
  pp. 1--8, 2019.

\bibitem{wladdimiro2016disaster}
D.~Wladdimiro, P.~Gonzalez-Cantergiani, N.~Hidalgo, and E.~Rosas, ``Disaster
  management platform to support real-time analytics,'' in \emph{2016 3rd
  International Conference on Information and Communication Technologies for
  Disaster Management (ICT-DM)}.\hskip 1em plus 0.5em minus 0.4em\relax IEEE,
  2016, pp. 1--8.

\bibitem{imran2016twitter}
M.~Imran, P.~Mitra, and C.~Castillo, ``Twitter as a lifeline: Human-annotated
  twitter corpora for nlp of crisis-related messages,'' \emph{arXiv preprint
  arXiv:1605.05894}, 2016.

\bibitem{nagar2014case}
R.~Nagar, Q.~Yuan, C.~C. Freifeld, M.~Santillana, A.~Nojima, R.~Chunara, and
  J.~S. Brownstein, ``A case study of the new york city 2012-2013 influenza
  season with daily geocoded twitter data from temporal and spatiotemporal
  perspectives,'' \emph{Journal of medical Internet research}, vol.~16, no.~10,
  p. e236, 2014.

\bibitem{szomszor2010swineflu}
M.~Szomszor, P.~Kostkova, and E.~De~Quincey, ``\# swineflu: Twitter predicts
  swine flu outbreak in 2009,'' in \emph{International conference on electronic
  healthcare}.\hskip 1em plus 0.5em minus 0.4em\relax Springer, 2010, pp.
  18--26.

\bibitem{odlum2015can}
M.~Odlum and S.~Yoon, ``What can we learn about the ebola outbreak from
  tweets?'' \emph{American journal of infection control}, vol.~43, no.~6, pp.
  563--571, 2015.

\bibitem{dredze2016zika}
M.~Dredze, D.~A. Broniatowski, and K.~M. Hilyard, ``Zika vaccine
  misconceptions: A social media analysis,'' \emph{Vaccine}, vol.~34, no.~30,
  p. 3441, 2016.

\bibitem{ordun2020exploratory}
C.~Ordun, S.~Purushotham, and E.~Raff, ``Exploratory analysis of covid-19
  tweets using topic modeling, umap, and digraphs,'' \emph{arXiv preprint
  arXiv:2005.03082}, 2020.

\bibitem{abd2020top}
A.~Abd-Alrazaq, D.~Alhuwail, M.~Househ, M.~Hamdi, and Z.~Shah, ``Top concerns
  of tweeters during the covid-19 pandemic: infoveillance study,''
  \emph{Journal of medical Internet research}, vol.~22, no.~4, p. e19016, 2020.

\bibitem{singh2020first}
L.~Singh, S.~Bansal, L.~Bode, C.~Budak, G.~Chi, K.~Kawintiranon, C.~Padden,
  R.~Vanarsdall, E.~Vraga, and Y.~Wang, ``A first look at covid-19 information
  and misinformation sharing on twitter,'' \emph{arXiv preprint
  arXiv:2003.13907}, 2020.

\bibitem{kouzy2020coronavirus}
R.~Kouzy, J.~Abi~Jaoude, A.~Kraitem, M.~B. El~Alam, B.~Karam, E.~Adib,
  J.~Zarka, C.~Traboulsi, E.~W. Akl, and K.~Baddour, ``Coronavirus goes viral:
  quantifying the covid-19 misinformation epidemic on twitter,'' \emph{Cureus},
  vol.~12, no.~3, 2020.

\bibitem{781w-ef42-20}
\BIBentryALTinterwordspacing
R.~Lamsal, ``Coronavirus (covid-19) tweets dataset,'' 2020. [Online].
  Available: \url{http://dx.doi.org/10.21227/781w-ef42}
\BIBentrySTDinterwordspacing

\bibitem{banda2020large}
J.~M. Banda, R.~Tekumalla, G.~Wang, J.~Yu, T.~Liu, Y.~Ding, and G.~Chowell, ``A
  large-scale covid-19 twitter chatter dataset for open scientific research--an
  international collaboration,'' \emph{arXiv preprint arXiv:2004.03688}, 2020.

\bibitem{chen2020tracking}
E.~Chen, K.~Lerman, and E.~Ferrara, ``Tracking social media discourse about the
  covid-19 pandemic: Development of a public coronavirus twitter data set,''
  \emph{JMIR Public Health and Surveillance}, vol.~6, no.~2, p. e19273, 2020.

\bibitem{kabir2020coronavis}
M.~Kabir, S.~Madria \emph{et~al.}, ``Coronavis: A real-time covid-19 tweets
  analyzer,'' \emph{arXiv preprint arXiv:2004.13932}, 2020.

\bibitem{sievert2017plotly}
C.~Sievert, C.~Parmer, T.~Hocking, S.~Chamberlain, K.~Ram, M.~Corvellec, and
  P.~Despouy, ``plotly: Create interactive web graphics via ‘plotly. js’,''
  \emph{R package version}, vol.~4, no.~1, p. 110, 2017.

\bibitem{loper2002nltk}
E.~Loper and S.~Bird, ``Nltk: the natural language toolkit,'' \emph{arXiv
  preprint cs/0205028}, 2002.

\bibitem{ramos2003using}
J.~Ramos \emph{et~al.}, ``Using tf-idf to determine word relevance in document
  queries,'' in \emph{Proceedings of the first instructional conference on
  machine learning}, vol. 242.\hskip 1em plus 0.5em minus 0.4em\relax New
  Jersey, USA, 2003, pp. 133--142.

\end{thebibliography}

\vspace{12pt}

\end{document}